\documentstyle[12pt]{article}
\newcommand{\esp}{\vspace{.1cm}}
\newcommand{\hsp}{\hspace{1cm}}

\newcommand{\smp}{\hspace{+2.5cm}}
\newcommand{\smn}{\hspace{-1cm}}
\newcommand{\ba}{\begin{array}}
\newcommand{\ea}{\end{array}}
\newcommand{\bc}{\begin{center}}
\newcommand{\ec}{\end{center}}
\newcommand{\be}{\begin{equation}}
\newcommand{\ee}{\end{equation}}
\newcommand{\bea}{\begin{eqnarray}}
\newcommand{\eea}{\end{eqnarray}}
\newcommand{\bdm}{\begin{displaymath}}
\newcommand{\edm}{\end{displaymath}}
\newcommand{\beas}{\begin{eqnarray*}}
\newcommand{\eeas}{\end{eqnarray*}}
\begin{document}
\begin{center}
{\Large \bf
                 Hyperspherical Adiabatic Formalism \\
                    of the Boltzmann Third Virial

}
\bigskip
{\large Sigurd Larsen}
\date{}
\smallskip

{\it
Physics Department, Temple University, Philadelphia Pa 19122, U.S.A.
}
\end{center}

\begin{abstract}
   First, we show that, if there are no bound states, we can express
   the q.m. third cluster - involving 3 and fewer particles in
   Statistical Mechanics - as a formula involving adiabatic eigenphase
   shifts. This is for Boltzmann statistics.

   From this q.m. formulation, in the case of purely repulsive forces,
   we recover, as $\hbar$ goes to $0$, the classical expressions for the
   cluster.

   We then discuss difficulties which arise in the presence of 2-body
   bound states and present a tentative formula involving eigenphase
   shifts and the 2 and 3 body bound state energies. We emphasize that
   important difficulties have not been resolved.
\end{abstract}
\noindent
\section*{Statistical Mechanics}
In equilibrium Statistical Mechanics ALL wisdom derives from the partition
function!
Here, we need the logarithm of the Grand Partition function ${\cal Q}$:
\beas
\ln {\cal Q} & = & z \, \; Tr(e^{-{\beta T_1}}) \\
      & + & z^2 \, [Tr(e^{-{\beta H_2}}) - \frac{1}{2}(Tr(e^{-{\beta T_1}}))
      ^2] \\
      & + & z^3 \, [Tr(e^{-{\beta H_3}}) - Tr(e^{-{\beta T_1}})
      Tr(e^{-{\beta H_2}})
  + \frac{1}{3}(Tr(e^{-{\beta T_1}})^3] \\
      & + & \cdots
\eeas
\noindent
which when divided by $V$, gives coefficients which are independent of the
volume, when the latter becomes large; we call them $b_l$.
The fugacity $z$ equals $\exp(\mu/\kappa T)$, where $\mu$ is the Gibbs
function per particle, $\kappa$ is Boltzmann's constant and $T$ is
the temperature; $\beta = 1/\kappa T$.
\esp
We can then write for the pressure and the density
\bdm
p/\kappa T = (1/V) \ln {\cal Q} = \sum_l b_l \, z^l
\edm
\bdm
N/V = \rho = \sum_l l \, b_l \, z^l
\edm
The fugacity can then be eliminated to give the pressure in terms
of the density.
\bdm
p/kT = \rho + \cdots
\edm
The coefficients of the second and higher powers are called the
virial coefficients.
\subsection*{Crucial Step}
For this work we extract the Boltzmann part of the traces: we write
\bdm
Tr(e^{- \beta H_n}) = \frac{1}{n!} Trace^{B}(e^{-\beta H_n}) +
Exchange \; Terms
\edm
We can then write for the Boltzmann $b_3$:
\bdm
b_3 = (3!V)^{-1}\, Trace^B[(e^{-\beta H_3} - e^{-\beta T_3})
- 3 \, (e^{-\beta(H_2 + T_1)} - e^{-\beta T_3})]
\edm
where I have made use of the Boltzmann statistics to express the answer in
terms of 3-body traces.
\noindent
\section*{Adiabatic Preliminaries}

  For the 3 particles of equal masses, in three dimensions, we first
introduce center of mass and Jacobi coordinates. We define
\bdm
 \vec{\eta}  =  (\frac{1}{2})^{1/2}(\vec{r}_{1} - \vec{r}_{2})
\hsp
 \vec{\xi}  =  (\frac{2}{3})^{1/2}(\frac{\vec{r}_{1} + \vec{r}_{2}}{2}
- \vec{r}_{3})
\hsp \vec{R}  =  \, \; \frac{1}{3} \; (\vec{r}_{1} + \vec{r}_{2} +
\vec{r}_{3})
\edm
where, of course, the $\vec{r}_{i}$ give us the locations of the 3 particles.
This is a canonical transformation and insures that in the kinetic energy
there are no cross terms.

The variables $\vec{\xi}$ and $\vec{\eta}$ are
involved separately in the Laplacians and we may consider them as acting in
different spaces. We introduce a higher dimensional vector $\vec{\rho} =
(\ba{c}\vec{\xi} \\ \vec{\eta}\ea)$ and express it in a hyperspherical
coordinate system ($\rho$ and the set of angles $\Omega$).
If we  factor a term of $\rho^{5/2}$ from the solution of the relative
Schr\"{o}dinger equation, i.e. we let $\psi =
\phi/\rho^{5/2}$, we are lead to:
\bdm
\left[ - \frac{\partial^2}{\partial \rho^2} + H_{\rho} - \frac{2 m E}{\hbar^2}
\right] \phi(\rho,\Omega) = 0
\edm
where
\bdm
H_{\rho} = - \frac{1}{\rho^2} \left[ \nabla^2_\Omega - \frac{15}{4} \right] +
\frac{2 m}{\hbar^2} V(\rho, \Omega)
\edm
and $m$ is the mass of each particle, $E$ is the relative energy in the
center of mass. \\$\nabla^2_{\Omega}$ is the purely angular part of the
Laplacian.
We now introduce the adiabatic basis, which consists
of the eigenfunctions of part of the Hamiltonian: the angular part of the
kinetic energy and the potential.
\bdm
H_{\rho} B_{\ell}(\rho,\Omega) = \Lambda_{\ell}(\rho) B_{\ell}(\rho,\Omega),
\edm
where $\ell$ enumerates the solutions.

   Using this adiabatic basis, we can now rewrite the
Schr\"{o}dinger equation as a system of coupled
ordinary differential equations.
\noindent
We write
\bdm \phi(\rho,\Omega) = \sum_{\ell^{\prime}} B_{\ell^{\prime}}(\rho,\Omega)
\tilde{\phi}_{\ell^{\prime}} (\rho)
\edm
and obtain the set of coupled equations
\beas
 ( \frac{d^{2}}{d\rho^{2}} - \Lambda_{\ell}(\rho) & + & k^{2} )
\tilde{\phi}_{\ell}(\rho) + 2\sum_{\ell^{\prime}} C_{\ell,\ell^{\prime}}\,
\frac{d}
{d\rho}\tilde{\phi}_{\ell^{\prime}}(\rho) \\
& + & \sum_{\ell^{\prime}}D_{\ell,\ell^{\prime}}\,
\tilde{\phi}_{\ell^{\prime}}(\rho) = 0 ,
\eeas
where
$k^{2}$ is the
relative energy multiplied by $2m/\hbar^{2}$ and we defined:
\beas
C_{\ell,\ell^{\prime}}(\rho) & = & \int d\Omega
\,B^{\ast}_{\ell}
(\Omega,\rho)\frac{\partial}{\partial\rho}B_{\ell^{\prime}}(\Omega,\rho) \\
D_{\ell,\ell^{\prime}}(\rho) & = & \int d\Omega \,
B^{\ast}_{\ell}
(\Omega,\rho)\frac{\partial^{2}}{\partial\rho^{2}}B_{\ell^{\prime}}
(\Omega,\rho) .
\eeas
We note that
\bdm
 D_{\ell,\ell^{\prime}} = \frac{d}{d\rho}\left(C_{\ell,\ell^{\prime}}\right)
 + \left(C^{2}\right)_{\ell,\ell^{\prime}} .
\edm
\section*{The Phase Shift Formula}
When there are no bound states,
we may
write
\bdm
Tr^{B}(e^{-\beta H_3}) = \int \!\! d\vec{\rho} \int \!\! dk \sum_{i}
\psi^{i}(k,\vec{\rho})  (\psi^{i}(k,\vec{\rho}))^{\ast} \;
e^{-\beta (\frac{\hbar^2}{2 m}k^2)}
\edm
where we have introduced a complete set of continuum eigenfunctions.
Expanding in the adiabatic basis, we obtain
\bdm
Tr^{B}(e^{-\beta H_3}) = \int \!\! d{\rho} \int
\!\! dk \sum_{i,\ell} |\tilde{\phi}^{i}_{\ell}(k,\rho)|^2 \;
e^{-\beta (\frac{\hbar^2} {2 m}k^2)} ,
\edm
where we note that we have integrated over the angles and taken advantage
of the orthogonality of our $B_{l}$'s. We integrate from $0$ to $\infty$.

We now return to our expression for $b_3$ and proceed as above, but
drop the tildas, to obtain:
\bdm
 \frac{3^{1/2}}{2 \lambda_{T}^3} \int \!\! dk\, e^{-\beta E_{k}}
\int \!\! d{\rho}
\sum_{i,\ell}[ (|{\phi}^{i}_{\ell}|^2 - |{\phi}^{i}_{\ell,0}|^2 )
- 3( |\bar{\phi}^{i}_{\ell}|^2  -  |\bar{\phi}^{i}_{\ell,0}|^2 )] ,
\edm
where we have evaluated the trace corresponding to the center of mass.
The amplitudes ${\phi}^{i}_{\ell}$ correspond to
$H_3$, $\bar{\phi}^{i}_{\ell}$ to $H_2 + T_1$ and
amplitudes with a zero belong to the free particles. The thermal wavelength
$\lambda_{T}$ is defined as $h/\sqrt{2 \pi m \kappa T}$.

\noindent
We now make use of a trick to evaluate the $\rho$ integrals.
We first write
\bdm
 \int_{0}^{\rho_{max} }\!   \sum_{\ell}  |\phi^{i}_{\ell}(k,\rho)|^2 \;
d\rho =
\lim_{k^{\prime} \rightarrow k}  \int_{0}^{\rho_{max}} \!
 \sum_{\ell} \phi^{i}_{\ell}(k,\rho) \phi^{i}_{\ell}(k^{\prime},\rho)
\, d\rho
\edm
and then, and there is the trick,
\beas
 \int_{0}^{\rho_{max} }\!   \sum_{\ell} \; (\!\! & \!\phi^{i}_{\ell}(k,\rho)&
\!\! \phi^{i}_{\ell}(k^{\prime},\rho))  d\rho = \\
\frac{1}{k^2 - (k^{\prime})^2} \sum_{\ell}\;[\! & \!\!\phi^{i}_{\ell}(k,\rho)&
\!\! \frac{d}
{d\rho}\phi^{i}_{\ell}(k^{\prime},\rho) - \phi^{i}_{\ell}(k^{\prime},\rho)
\frac{d}{d\rho}\phi^{i}_{\ell}(k,\rho)] ,
\eeas
evaluated at $\rho = \rho_{max}$. \\
\noindent
---------------------------------------------------------------- \\
I.e. our identity is:
\beas
&& \ \sum_{\ell}     \frac{d}{d\rho} \left[\phi_{\ell}(k^{\prime})
\frac{d}{d\rho}
\phi_{\ell}(k) - \phi_{\ell}(k)\frac{d}{d\rho}\phi_{\ell}(k^{\prime})\right] \\
&& + \left(k^2 - (k^{\prime})^2\right)\sum_{\ell} \phi_{\ell}(k) \
\phi_{\ell}(k^{\prime}) \\
&& + 2 \ \sum_{\ell,{\ell}^{\prime}} \frac{d}{d\rho}
\left[ \phi_{\ell}(k^{\prime})
\ C_{\ell,{\ell}^{\prime}} \ \phi_{{\ell}^{\prime}}(k) \right] = 0
\eeas
and we integrate with respect to $\rho$. Using then the fact that $\phi$
goes to zero, as $\rho$ itself goes to zero, and that C decreases fast
enough for $\rho$ large, we are left with the expression
displayed earlier (that of our `trick'). \\
---------------------------------------------------------------- \\
\noindent
We now put in the asymptotic form of our solutions, oscillatory solutions
valid for $\rho_{max}$ large, and use l'Hospital's rule to take the
limit as $k^{\prime}  \rightarrow k$. \\
\noindent
The solutions are:
\bdm
\phi^{i}_{\ell} \rightarrow (k\rho)^{1/2} {\cal C}_{\ell,i} \,
[\cos \delta_{i}\: J_{K+2}(k\rho) - \sin \delta_{i}\: N_{K+2}(k\rho)]
\edm
where the order $K$ is one of the quantities specified by $\ell$.
Inserting this into our integrals we find that
\bdm
\sum_{\ell} \int_{0}^{\rho_{max}}\! |\phi^{i}_{\ell}(k)|^2 \:  d\rho
\rightarrow
\frac{1}{\pi} \frac{d}{dk} \delta^{i}(k)
+ \frac{1}{\pi} \rho_{max} \: + \: osc. \: terms
\edm
and, thus, that
\bdm
\int_{0}^{\rho_{max}}\! (| \phi^{i}_{\ell}(k)|^{2}
 -  |\phi^{i}_{\ell,0}(k)|^2) \; d\rho  \rightarrow
\frac{1}{\pi} \frac{d}{dk} \delta^{i}(k) \: + \: osc. \: terms
\edm
We let $\rho_{max}$ go to infinity, and the oscillating terms
- of the form $\sin(2 k \rho_{max} + \cdots)$ - will
not contribute to the subsequent integration over $k$.
A partial integration now gives us our basic formula.

\bdm
b_3^{Boltz} = \frac{3^{1/2}}{(2 \pi)^{2}\lambda_{T}}\int_{0}^{\infty}
\! dk \; k \, G(k) \;
e^{-\beta \frac{\hbar^2}{2 m} k^{2}}
\edm
where
\bdm
G(k) = \sum_{i}\: [ \delta_{i}(k) - 3 \, \bar{\delta}_{i}(k)]
\edm
The first $\delta$ arises from comparing three interacting particles
with three free particles. The second $\bar{\delta}$  arises when a 3-body
system, where only two particles are interacting (one particle being a
spectator), is compared to three free particles.

\section*{Classical Limit}
The idea behind our WKB treatment of our equations, is to argue that when
the potentials change slowly - within oscillations of the solutions -
then the adiabatic eigenfunctions will also change slowly and we can neglect
their derivatives. Thus we will obtain {\bf uncoupled}
equations with effective potentials (the eigenpotentials
$\Lambda_{\ell}(\rho)$).
We then proceed with these in a more or less
conventional WKB fashion.  Let us assume, here, one turning point $\rho_{0}$.

The phases can now be obtained by considering simplified forms of the
asymptotic solutions for the $\phi's$. Let us denote them as $\phi_{\nu}$.
The phases will then be
\bdm
\delta_{\nu} \sim (K+2)\frac{\pi}{2} - k\rho_{0} +
\int_{\rho_{0}}^{\infty}[\sqrt{k^2 \! - \! \Lambda_{\nu} \!
- \! \frac{1}{4\rho^2}} \! - \! k] d\rho \\
\edm
Inserting our expression for $\delta_{\nu}$ into
$ \int_{0}^{\infty} dk\: k \: \delta_{\nu}(k) \:
\exp(- {{\lambda_{T}^{2}} k^{2}/{4 \pi}} )$
and interchanging the order of integration ($\rho$ and $k$) we obtain:
\bdm
\frac{2(\pi^2)}{\lambda_{T}^{3}}\int_{0}^{\infty}d\rho
\{\exp[-\frac{\lambda_{T}^2}{4 \pi}(\Lambda_{\nu} + \frac{1}{4\rho^2})]
- \exp[-\frac{\lambda_{T}^2}{4 \pi}\frac{(K+2)^2}{\rho^2}]\} .
\edm
Summing now over $\nu$, we can rewrite the exponentials as traces:
\beas
\sum_{\nu}\!\!\!\! &\{&\!\!\!\!\exp[-\frac{\lambda_{T}^2}{4 \pi}\,
(\Lambda_{\nu}\! +\! \frac{1}{4\rho^2})]
- \exp[-\frac{\lambda_{T}^2}{4 \pi}\,\frac{(K+2)^2}{\rho^2}]\} \\
 &=& \mbox{Trace}^{R}\,
\{\exp[-\frac{\lambda_{T}^2}{4 \pi}\,(\Lambda(\rho) + \frac{1}{4 \rho^2})] -
\exp[-\frac{\lambda_{T}^2}{4 \pi}\,\frac{{\cal K}^2\! +\!
\frac{1}{4}}{\rho^2}] \}
\eeas
where $\Lambda$ is the operator (matrix) which yields the diagonal
elements $\Lambda_{\nu}$ and ${\cal K}^2$ the operator which yields
the eigenvalue when the interaction is turned off (and therefore
takes on the diagonal values $(K+2)^2 - \frac{1}{4}$,
associated with the hyperspherical harmonic of order $K$).
The trace is restricted so as not to involve $\rho$.

\noindent
In another key step, we switch to a hyperspherical basis.
We note that $\Lambda$ is related to
$(2 m/\hbar^2) \, V  + {\cal K}^2/\rho^2$ by a similarity transformation
and an orthogonal matrix $U$. Substituting in the trace, we lose the $U$
and obtain
\bdm
\mbox{Tr}^{R}\,[\exp(-\beta V -\frac{\lambda_{T}^2}{4 \pi}\,
\frac{{\cal K}^2 + \frac{1}{4}}{\rho^2}) -
\exp(-\frac{\lambda_{T}^2}{4 \pi}\,
\frac{{\cal K}^2 + \frac{1}{4}}{\rho^2}) ]
\edm
We write the exponential as a product of 2 exponentials, disregarding
higher order terms in $\hbar$. Introducing eigenkets and eigenbras which
depend on the hyperspherical angles, we write the trace as:
\bdm
\int d\Omega <\Omega|\exp(-\frac{\lambda_{T}^2}{4 \pi}\,
\frac{{\cal K}^2 + \frac{1}{4}}{\rho^2})|\Omega>\{
\exp[-\beta V(\vec{\rho})] -1\}
\edm

The matrix element above can be evaluated and, to leading order in a
Euler McLaurin expansion, yields $\rho^5/\lambda_{T}^5$.
For the phase shifts of type $\delta_{\nu}$, associated with the fully
interacting 3 particles, $V$ equals $V(12) + V(13) + V(23)$ and we obtain
as its contribution to $b_3^{Boltz}$:
\bdm
\frac{3^{1/2}}{2 \lambda_{T}^9}
\int\! d\vec{\xi}\, d\vec{\eta}\,
( \exp[ - \beta ( V(12)\! + \! V(13)\! +\! V(23) ) ]\! -\! 1 )
\edm
The expression above, derived solely from the contribution of the
$\delta$'s, diverges for infinite volume. However, including the terms
in $\bar{\delta}$, associated with the pairs 12, 13 and 23 provides
a convergent answer.
The complete result for $b_3^{Boltz}$ divided by $b_1^3$, where
$b_1 = \lambda_{T}$, equals
\beas
\smn &\mbox{}& \frac{1}{3!V}
\int\! d\vec{r}_1 \, d\vec{r}_2 \, d\vec{r}_3 \,
\{ \exp[ - \beta ( V(12)\! + \! V(13)\! +\! V(23) )] \\
\smn &\mbox{}&- \exp[-\beta V(12)] - \exp[-\beta V(13)] - \exp[-\beta V(23)]
+ 2 \}
\eeas
where I have integrated over $\vec{R}$ the center of mass coordinate,
divided by V, and changed to the coordinates $\vec{r}_1,\, \vec{r}_2 \,
\mbox{and} \,  \vec{r}_3$.
The result is the classical expression with all the correct factors.
\section*{Bound States}
If there are bound states, the major change in the eigenpotentials
is that for some of these potentials, instead of going to zero at large
distances (large $\rho$), there appears a negative `plateau'. I.e.
the eigenpotential (up to some contribution in $1/\rho^2$), becomes flat
and negative. This is the indication that asymptotically the physical
system consists of a 2-body bound state and a free particle.
The eigenpotential may also `support' one or more 3-body bound states.

\noindent
The eigenfunction expansion of the trace associated with $H_3$, will
read:
\beas
\smn &\mbox{}& \sum_{m} e^{-\beta E_{3,m}} +
\sum_{i} \int_{0}^{\infty} \!\! dk \!
\int \!\! d\vec{\rho}\,  \psi^{i}(k,\vec{\rho})\,
(\psi^{i}(k,\vec{\rho}))^{\ast} \;
e^{-\beta (\frac{\hbar^2}{2 m}k^2)} \\
\smn &\mbox{}&
\smp  +  \sum_{i}\! \int_{0}^{q_i}  \!\! dq \!
 \int \!\! d\vec{\rho} \,
 \psi^{i}(q,\vec{\rho}) \, (\psi^{i}(q,\vec{\rho}))^{\ast} \;
e^{-\beta (\frac{\hbar^2}{2 m}q^2 -\epsilon_{2,i})}
\eeas
The $q$'s are defined by $k^2 = q^2 - \epsilon_{2,i}$, where
$\epsilon_{2,i}$ is the binding energy of the corresponding bound state.
The limit $q_{i}$ equals  $\sqrt{\frac{2 m}{\hbar^2}\epsilon_{2,i}}$.
The new continuum term represents solutions which are still
oscillatory for negative
energies (above that of the respective bound states).

Assume, now, that we have 1 bound state, and introduce amplitudes.
The asymptotic behaviour will be as follows. \\
For $E > 0$.
\bdm
\phi^{i}_{\ell}(\rho) \rightarrow (k\rho)^{1/2}{\cal C}_{\ell,i}\,
[\cos \delta_{i} \, J_{K_{\ell}+2}(k\rho) - \sin \delta_{i} \,
N_{K_{\ell}+2}(k\rho)]
\edm
\bdm
\phi^{i}_{\ell_{0}}(\rho) \rightarrow (k\rho)^{1/2}{\cal C}_{\ell_{0},i}\,
[\cos \delta_{i} \, J_{K_{\ell_{0}}+2}(q\rho) - \sin \delta_{i} \,
N_{K_{\ell_{0}}+2}(q\rho)]
\edm
Using our procedure as before we obtain for the integral over $\rho$:
\bdm
\frac{1}{\pi}\frac{d}{dk} \delta_{i} + \frac{\rho_{max}}{\pi}(
\sum_{\ell \neq \ell_{0}} |{\cal C}_{\ell,i}|^2 + |{\cal C}_{\ell_{0},i}|^2
\frac{k}{q})
\edm
For $E < 0$.
\bdm
\phi^{i}_{\ell_{0}}(\rho) \rightarrow (q\rho)^{1/2}
[\cos \delta_{i} \, J_{K_{\ell_{0}}+2}(q\rho) - \sin \delta_{i} \,
N_{K_{\ell_{0}}+2}(q\rho)]
\edm
which then yields
\bdm
\frac{1}{\pi}\frac{d}{dk} \delta_{i} + \frac{\rho_{max}}{\pi}
\edm
{\bf The problem} is that I can no longer eliminate the $\rho_{max}$ term
by subtracting the contribution of the free particle term; i.e.
using the $\rho_{max}$ from $T_3$ to cancel the $\rho_{max}$ from $H_3$.
{\bf All is not lost} however, as we saw (for example in the terms
arising in the classical limit) that all the terms of the cluster ($b_3$)
are needed to obtain a volume independent and convergent result.
The obvious terms to examine are the ones associated with $H_2 + T_1$, which
also have amplitudes that correspond to (2-body) bound states.
%
I
have not been able, to date, to prove that all the coefficients
are such that
the final coefficient of $\rho_{max}$ is zero.

{\bf If we were ...} to assume that the terms in $\rho_{max}$ {\bf do}
indeed cancel, then we can write the following formula for the
complete trace.
\beas
\smn &\mbox{}&
Trace^B[(e^{-\beta H_3} - e^{-\beta T_3})
- 3 \, (e^{-\beta(H_2 + T_1)} - e^{-\beta T_3})] = \\
\esp \esp \esp
\smn &\mbox{}& \sum_{m} e^{-\beta E_{3,m}} +
\frac{1}{\pi} \sum_{i} \int_{0}^{\infty} \!\! dk \!
\, \frac{d}{dk} [\delta_{i}(k) - 3 \bar{\delta}_{i}(k)]
e^{-\beta (\frac{\hbar^2}{2 m}k^2)} \\
\smn &\mbox{}&
\smp  +  \frac{1}{\pi} \sum_{i}\! e^{\beta \epsilon_{i}}
\int_{0}^{q_i}  \!\! dq \!
\, \frac{d}{dq} [\delta_{i}(q) - 3 \bar{\delta}_{i}(q)]
e^{-\beta (\frac{\hbar^2}{2 m}q^2)}
\eeas

\end{document}